\documentclass[aps,preprint,draft]{revtex4}
\input epsf
\begin{document}

\title{
Variational description of the dimensional cross-over
in the array of coupled one-dimensional conductors
}
\author{A.V. Rozhkov}

\affiliation{
Center~for~Materials~Theory,
Department~of~Physics~and~Astronomy, Rutgers~University,
136~Frelinghuysen~Road,
Piscataway, NJ~08854,~USA
}


\begin{abstract}
Variational wave function is proposed to describe electronic properties
of an array of one-dimensional conductors coupled by transverse hopping
and interaction.  For weak or intermediate in-chain interaction the wave 
function has the following structure: Tomonaga-Luttinger bosons with 
momentum higher then some 
variational quantity $\tilde\Lambda$ are in their ground state while other
bosons (with $|k|<\tilde\Lambda$) form kinks -- fermion-like excitations of
the Tomonaga-Luttinger boson field. Nature of the ground state for this 
quasiparticles can be determined by solving three dimensional 
effective hamiltonian. Since the anisotropy of the effective hamiltonian
is small the use of the mean field theory is justified. For repulsive
interaction possible phases are density wave and $p$-wave
superconductivity. Our method allows us to calculate the low-energy part
of different electronic Green's functions. In order to do that it is enough
to apply standard perturbation theory technique to the effective
hamiltonian. When the
in-chain interaction is strong $\tilde\Lambda$ vanishes and no fermionic
excitation is present in the system. In this regime the dynamics is
described by transversally coupled Tomonaga-Luttinger bosons.
\end{abstract}

\maketitle
\hfill

\section{Introduction}

The adequate description of quasi-one-dimensional (Q1D) conductors remains 
an unresolved theoretical challenge. Experimentally, at low temperature 
such systems either three dimensional anisotropic Fermi liquids or they 
freeze into a three dimensional phase with broken symmetry\cite{book}. 
At high temperature their transport properties show many unusual 
features generally attributed to the one-dimensional electron anisotropy. 
This cross-over from 1D to 3D is the core problem of Q1D physics.

It is possible to look at the issue of the dimensional cross-over from
another angle. At high temperature the proper elementary excitations of the
system are Tomonaga-Luttinger (TL) bosons. When the temperature is low and
the interaction is weak enough the elementary excitations are fermions.
Therefore, to describe the system at different energy scales one needs 
to explain how high-energy bosons `cross over' into low-energy fermions.
Obviously, this is a non-trivial task.

In this paper we develop a variational approach which accomplishes this
goal. To explain the structure of the variational wave function let us first
consider
a one-dimensional conductor described by TL hamiltonian. The ground state
of this system is the ground state of TL bosons with all momenta $k$. Let's
turn the transverse hopping on and couple $N_\perp$ of these conductors 
into 3D array. In this situation the system will attempt to lower its ground
state energy even further by taking advantage of the transverse hopping
energy. However, in order to participate in
hopping the bosons have to form many-body fermion-like excitations which
have finite overlap with the physical fermion.

To accommodate for possibility of having two types of excitations, bosonic
and fermionic, we device our variational state in the following fashion. We
introduce intermediate cut-off $\tilde\Lambda<\Lambda$,
where $\Lambda$ is the cut-off of the 1D hamiltonian. All TL bosons whose
energy and momenta are high ($|k|>\tilde\Lambda$) remain in their ground 
states. The small momenta bosons ($|k|<\tilde\Lambda$) form fermion-like
excitations which are delocalized in transverse direction. To distinguish
between the physical electrons and these fermionic excitations we will
refer to the latter as quasiparticles. In other words, the wave function
can be factorized into two parts. High-energy part corresponds to the
ground state of $|k|>\tilde\Lambda$ TL bosons, low-energy part corresponds
to the 3D anisotropic Fermi liquid composed of the quasiparticles. 

The variational energy is minimized by adjusting $\tilde \Lambda$. The
energy of quasiparticle transverse hopping is decreasing function of
$\tilde \Lambda$. At the same time, the in-chain energy grows when $\tilde
\Lambda$ grows. The trade-off between the transverse kinetic energy
and the in-chain potential energy determines the value of $\tilde \Lambda$.

If the optimal value of $\tilde \Lambda$ is non-zero the low-energy
excitations of the system are the quasiparticles.
Properties of the fermionic quasiparticle state depend on quasiparticle
effective hamiltonian. It arises naturally after
high-energy bosons are `integrated out'. In this effective hamiltonian the
anisotropy is insignificant. Standard many-body techniques such as 
perturbation theory and mean field theory can be used to calculate
Green's functions and map out the quasiparticle phase diagram.
Since the physical electron and the quasiparticle have finite
overlap there is a direct correspondence between the broken symmetry phases
of the effective hamiltonian and the physical system. We will show that
possible phases for spinless Q1D electrons with repulsion are the charge 
density wave (CDW) and the superconductivity with the Cooper pairs formed 
of the electrons on neighboring 1D chains. 

As the in-chain interaction grows parameter $\tilde\Lambda$ approaches zero.
When $\tilde\Lambda$ vanish the fermionic excitations cease to exist. 
The system is described by 3D TL boson state. In such a regime the ground
state is CDW.

Our approach allows us to obtain some new analytical results. With the help
of the method it is possible to derive a formula for quasiparticle damping
near Fermi surface. Also, we evaluate transition temperatures for CDW and 
superconductivity. The knowledge of these temperatures allows us to map out
the phase diagram of our system. Although, these quantities have been
obtained using different numerical techniques \cite{arrigoni,dup} the 
analytical expressions had not been reported.

The paper is organized as follows. In Section II we determine
$\tilde\Lambda$ and derive the effective hamiltonian for the fermions.
Section III contains the evaluation of the single-particle Green's function.
Different phases of the effective hamiltonian (and the physical system) are
mapped in Section IV. The regime where $\tilde \Lambda = 0$ is discussed in
Section V.  We give our conclusions in Section VI.

\section{Variational procedure}

We start our analysis by writing down the hamiltonian for the array of
coupled 1D conductors:
\begin{eqnarray}
&&H = \int_0^L dx {\cal H},\label{H}\\
&&{\cal H} = \sum_i {\cal H}_i^{\rm 1d} + \sum_{i,j} {\cal H}_{ij}^\perp,\\
&&{\cal H}_i^{\rm 1d} = {\rm i} v_{\rm F} \left( \psi^\dagger_{{\rm L}i}
\nabla\psi^{\vphantom{\dagger}}_{{\rm L}i} - \psi^\dagger_{{\rm R}i}
\nabla\psi^{\vphantom{\dagger}}_{{\rm R}i}\right) +
g\psi_{{\rm L}i}^\dagger \psi_{{\rm L}i}^{\vphantom{\dagger}}
\psi_{{\rm R}i}^\dagger \psi_{{\rm R}i}^{\vphantom{\dagger}},\label{H1d}\\
&&{\cal H}_{ij}^\perp = - t(i-j) \sum_{p={\rm L,R}}
\left(\psi^\dagger_{pi}\psi^{\vphantom{\dagger}}_{pj} + {\rm h.c.}\right)
+g_{2k_{\rm F}}(i-j) \left( \psi^\dagger_{{\rm L}i}
\psi^{\vphantom{\dagger}}_{{\rm R}i} \psi^\dagger_{{\rm R}j}
\psi^{\vphantom{\dagger}}_{{\rm L}j} + {\rm h.c.}\right)
\label{perp}\\
&&+g_{0}(i-j) \left(\psi^\dagger_{{\rm L}i}
\psi^{\vphantom{\dagger}}_{{\rm L}i} + \psi^\dagger_{{\rm R}i}
\psi^{\vphantom{\dagger}}_{{\rm R}i} \right)
\left(\psi^\dagger_{{\rm L}j} \psi^{\vphantom{\dagger}}_{{\rm L}j} + 
\psi^\dagger_{{\rm R}j} \psi^{\vphantom{\dagger}}_{{\rm R}j} \right),
\nonumber
\end{eqnarray}
with the real-space cut-off $a=\pi/\Lambda$. The fermionic field
$\psi_{pi}^\dagger$ creates physical electron with the chirality $p={\rm
L}(+)$ or $p={\rm R}(-)$ on chain $i$. Transverse interaction constants
$g_{0}$ (forward scattering) and $g_{2k_{\rm F}}$ (exchange) are positive. 
The terms proportional to $g_0$ and $g_{2k_{\rm F}}$ account for the
Coulomb repulsion of the electrons on different chains. It is further
assumed that:
\begin{equation}
g > g_{0} > g_{2k_{\rm F}}.\label{int}
\end{equation} 
Now we use Abelian bosonization prescription \cite{boson}:
\begin{eqnarray}
\psi^\dagger_{p} (x) =
(2\pi a)^{-1/2} \eta_{p}{\rm e}^{{\rm i}\sqrt{2\pi}
\varphi_{p}(x)}=
(2\pi a)^{-1/2} \eta_{p}{\rm e}^{{\rm i}\sqrt{\pi} \left[
\Theta(x) + p\Phi(x) \right]},\label{bos}
\end{eqnarray}
to express the electron hamiltonian in terms of bosonic fields. In the 
above formula $\eta_{p}$ are Klein factors, $\Theta$ is the TL boson field, 
$\Phi$ is the dual field. The bosonized one-chain hamiltonian is:
\begin{equation}
{\cal H}^{\rm 1d} \left[\Theta,\Phi\right]= 
\frac{v_{\rm F}}{2} \left( \colon\left(\nabla \Theta \right)^2\colon + 
\colon\left(\nabla \Phi \right)^2\colon \right) + 
\frac{g}{4\pi} \left(\colon\left(\nabla \Phi \right)^2\colon -
\colon\left(\nabla \Theta \right)^2\colon \right).
\end{equation}
The symbol $\colon\ldots\colon$ denotes normal ordering of TL boson operators 
with respect to non-interacting ($g=0$) ground state. 

Let us introduce our main variational parameter $\tilde \Lambda < \Lambda$
and use it to split TL boson fields
into fast ($\Lambda>||k_\|| - k_{\rm F}| > \tilde\Lambda$, subscript `$>$')
and slow ($||k_\|| - k_{\rm F}| < \tilde\Lambda$, subscript `$<$') modes:
\begin{eqnarray}
{\cal H}^{\rm 1d} \left[\Theta,\Phi\right] = {\cal H}^{\rm 1d}_< +
{\cal H}^{\rm 1d}_> = {\cal H}^{\rm 1d} \left[\Theta_<,\Phi_<\right] +
{\cal H}^{\rm 1d} \left[\Theta_>,\Phi_>\right].
\end{eqnarray}
We define the fermionic field $\Psi_p^\dagger(x)$ with the help of
equation (\ref{bos}) in which $a$ is substituted by
$\tilde a = \pi/\tilde\Lambda$ and $\Theta_<$ and $\Phi_<$ are placed
instead of $\Theta$ and $\Phi$. The field $\Psi$ is
our quasiparticle discussed in Introduction. Using this field
we re-fermionize ${\cal H}_<^{\rm 1d}$. The result is the same as 
(\ref{H1d}) with $\Psi_p$ instead of $\psi_p$. The transverse terms 
(\ref{perp}) can be easily re-written if one observe that the physical 
fermion is simply:
\begin{equation}
\psi_p^\dagger = \sqrt{\tilde a/a}\Psi_p^\dagger 
{\rm e}^{{\rm i}\sqrt{\pi} ( \Theta_> + p\Phi_>)},\label{physical}
\end{equation}
and that the fermionic and bosonic parts in this definition commute with
each other. Therefore, ${\cal H}_{ij}^\perp$ is equal to
\begin{eqnarray}
{\cal H}_{ij}^\perp&=&- (\tilde a/a) t(i-j) \sum_{p={\rm L,R}}
\left(\Psi^\dagger_{pi}\Psi^{\vphantom{\dagger}}_{pj} {\rm e}^{{\rm i}
\sqrt{\pi} ((\Theta_{>i} - \Theta_{>j}) + p(\Phi_{>i} - \Phi_{>j}))}
+ {\rm h.c.}\right)\\
&&+(\tilde a/a)^2 g_{2k_{\rm F}}(i-j) \left(\Psi^\dagger_{{\rm L}i}
\Psi^{\vphantom{\dagger}}_{{\rm R}i} \Psi^\dagger_{{\rm R}j}
\Psi^{\vphantom{\dagger}}_{{\rm L}j} {\rm e}^{{\rm i}\sqrt{4\pi}
(\Phi_{>i}-\Phi_{>j})} + {\rm h.c.}\right) \nonumber\\
&&+g_{0}(i-j) \left(\left(\Psi^\dagger_{{\rm L}i}
\Psi^{\vphantom{\dagger}}_{{\rm L}i} + \Psi^\dagger_{{\rm R}i}
\Psi^{\vphantom{\dagger}}_{{\rm R}i}\right)
\left(\Psi^\dagger_{{\rm L}j}
\Psi^{\vphantom{\dagger}}_{{\rm L}j} + \Psi^\dagger_{{\rm R}j}
\Psi^{\vphantom{\dagger}}_{{\rm R}j}\right)
+ \frac{1}{\pi}  \nabla \Phi_{>i} \nabla \Phi_{>j} \right).
\nonumber
\end{eqnarray}
Our variational wave function has the form:
\begin{equation}
\left|{\rm Var}\right> = \left| \left\{ \Psi_{pi} \right\} \right> \prod_j
\left| 0_{>j} \right> =
\left| \left\{ \Psi_{pi} \right\} \right> \prod_{j,k>\tilde \Lambda}
(2|k|/\pi {\cal K})^{1/2}
\exp \left\{-|k|\left|\Phi_{jk} \right|^2/ {\cal K} \right\}.
\label{vari}
\end{equation}
It is a product of some many-body state 
$\left| \left\{ \Psi_{pi} \right\} \right>$ composed of the quasiparticles
$\Psi_{pi}$ and the ground states $\left| 0_{>j} \right>$ of 
${\cal H}^{\rm 1d}[\Theta_{>j}, \Phi_{>j}]$.

Variational ground state energy is found by minimizing the expression:
\begin{eqnarray}
E^{\rm V}&=&N_\perp L \frac{\theta u}{4\pi} \left( {\tilde\Lambda}^2 - 
\Lambda^2 \right)
+\left<\left\{ \Psi_{pi} \right\}\right|\left(\int_0^L dx {\cal H}^{\rm eff} 
\right) \left|\left\{\Psi_{pi} \right\}\right>,\label{EV}\\
{\cal H}^{\rm eff}&=&\sum_i
{\rm i} v_{\rm F} \left( \Psi^\dagger_{{\rm L}i}
\partial_x \Psi^{\vphantom{\dagger}}_{{\rm L}i} - \Psi^\dagger_{{\rm R}i}
\partial_x \Psi^{\vphantom{\dagger}}_{{\rm R}i}\right) + g\Psi_{{\rm
L}i}^\dagger \Psi_{{\rm L}i}^{\vphantom{\dagger}}
\Psi_{{\rm R}i}^\dagger \Psi_{{\rm R}i}^{\vphantom{\dagger}}\label{Heff}\\
&-&\sum_{ij}  \sum_{p={\rm L,R}}{\tilde t} (i-j) 
\left(\Psi^\dagger_{pi}\Psi^{\vphantom{\dagger}}_{pj} + {\rm h.c.}\right)+
\sum_{ij}{\tilde g}_{2k_{\rm F}}(i-j)\left( \Psi^\dagger_{{\rm L}i}
\Psi^{\vphantom{\dagger}}_{{\rm R}i} \Psi^\dagger_{{\rm R}j}
\Psi^{\vphantom{\dagger}}_{{\rm L}j} + {\rm h.c.}\right)\nonumber\\
&+&g_{0}(i-j) \left(\Psi^\dagger_{{\rm L}i}
\Psi^{\vphantom{\dagger}}_{{\rm L}i} + \Psi^\dagger_{{\rm R}i}
\Psi^{\vphantom{\dagger}}_{{\rm R}i} \right)
\left(\Psi^\dagger_{{\rm L}j} \Psi^{\vphantom{\dagger}}_{{\rm L}j} +
\Psi^\dagger_{{\rm R}j} \Psi^{\vphantom{\dagger}}_{{\rm R}j}
\right),\nonumber \\
\tilde t&=&\zeta^{\theta} t,\quad
\tilde g_{2k_{\rm F}} = \zeta^{2{\cal K}-2} g_{2k_{\rm F}},\quad
\zeta=\tilde\Lambda/\Lambda.\label{eff_param}
\end{eqnarray}
The number of chains is $N_\perp$.  The TL liquid parameter 
${\cal K}$, the electron anomalous dimension $\theta$ and boson velocity
$u$ are defined in the usual way:
\begin{equation}
{\cal K} = \sqrt{\frac{2\pi v_{\rm F} - g}{2\pi v_{\rm F} + g}},\quad
\theta = \frac{1}{2} \left( {\cal K} + {\cal K}^{-1} - 2 \right),\quad
u= \frac{1}{2\pi}\sqrt{\left(2\pi v_{\rm F} - g\right)\left(2\pi v_{\rm F} 
+ g\right)}.
\end{equation}
The first term of (\ref{EV}) has purely one-dimensional origin. The second
term is the energy of the quasiparticle ground state.

Observe that the parameters of the effective hamiltonian
$\tilde t/\tilde \Lambda$ and $\tilde g_{2k_{\rm F}}$ are connected to the
corresponding bare parameters
as if they are subject to the renormalization group (RG) flow in the
vicinity of the TL fixed point. The explanation to this fact is quite
obvious: our method of deriving the effective hamiltonian is equivalent to
the tree level RG scaling near TL fixed point.

If the transversal interactions are small ($g_0$ and
$\tilde g_{2k_{\rm F}}$ both less then $\tilde t/\tilde \Lambda$) they can
be neglected.  In addition, we neglect corrections to the energy due to
spontaneous symmetry breaking. The latter assumption works when
$\theta \ll 1$. Its validity away from this point will be discussed at the
end of Section V.  Under these two conditions the expression (\ref{EV})
becomes:
\begin{equation}
E^{\rm V}/(LN_\perp) \approx \frac{\theta u}{4\pi} \left(\zeta^2-1\right)
\Lambda^2 -\frac{2}{\pi v_{\rm F}} \zeta^{2\theta} \sum_i 
\left(t(i)\right)^2.\label{EVs}
\end{equation}
This variational energy attains its minimum at 
\begin{eqnarray}
\zeta = \cases{ \left( 8{\bar t}^2/u v_{\rm F} 
\Lambda^2 \right)^{1/(2-2\theta)} & if $\theta<1$, \cr
0 & if $\theta>1$,\cr}\label{L8}\\
{\bar t}^2 = \sum_i \left(t(i)\right)^2.
\end{eqnarray}
We have to remember, however, that value of the numerical coefficient
in (\ref{L8}) is not accurate. This is due to the fact that the second term
in (\ref{EVs}) is calculated under assumption $\tilde t < \tilde \Lambda$.
When $\tilde\Lambda$ gets smaller the coefficient in front of this term
acquires some $\tilde\Lambda$ dependence. We neglect the corrections due to
this dependence since they are less singular (at small $\tilde \Lambda$)
then the second term of (\ref{EVs}). These corrections modify the result
for $\zeta$ quantitatively, therefore, it is more appropriate to
write
\begin{eqnarray}
\zeta \propto  \left(\frac{t}{\bar u\Lambda}
\right) ^{1/(1-\theta)},\\
\bar u = \sqrt{u v_{\rm F}}.
\end{eqnarray}
Using this formula it is easy to show that:
\begin{equation}
\tilde t \propto \bar u \tilde\Lambda \propto t \left( \frac{t}{\bar u
\Lambda} \right)^{\theta/(1-\theta)}.
\end{equation}
This means that for $\theta<1$ the effective transverse
hopping amplitude $\tilde t$ of the quasiparticle $\Psi$ is
of the same order as the quasiparticle longitudinal cut-off energy $v_{\rm F} 
\tilde\Lambda$. Therefore, due to small anisotropy, the hamiltonian for the
quasiparticles (\ref{Heff}) can be treated within the framework of usual
mean field theory and perturbation theory.

Our calculations, in agreement with renormalization group analysis 
\cite{bour,bourII}, show that for $\theta<1$ there is the cross-over 
energy scale $\tilde t$ above which the system is equivalent to a 
collection of decoupled chains while below the transverse hopping becomes
important.

Depending on the interaction and the anisotropy the region
$\theta < 1$ can be further split into two parts. The transverse hopping
contribution to the variational energy (second term of (\ref{EVs})) can be
re-written as follows:
\begin{equation}
\frac{2}{\pi v_{\rm F}} {\tilde t}^2 \propto \frac{2}{\pi v_{\rm F}}
t^2 \exp \left(- \frac{2\theta}{1-\theta} \log \left( 
\frac{\bar u \Lambda}{t} \right)\right).
\end{equation}
If the argument of the exponential function is small the exponential can be
replaced by the first few terms of the Taylor series. In such a situation 
the contribution of the in-chain interaction to the total energy, eq.
(\ref{EVs}), can be calculated perturbatively. One-dimensional effects are
virtually unobservable. This is the weak coupling regime.

When the anisotropy and the in-chain interaction are strong the exponential
cannot be approximated accurately by the low order Taylor expansion. The
system is in the intermediate coupling regime now. In order to obtain a
reliable answer in such a regime it is not enough to apply finite-order
perturbation theory. Our method converts the system of physical electrons
with intermediate coupling into the system of quasiparticles with weak
coupling. The latter can be studied by standard perturbation theory.

As a function of the bare transverse hopping amplitude $t$ the cross-over
from the weak coupling to the intermediate coupling occurs at:
\begin{equation}
t^* \propto \bar u \Lambda \exp \left( -\frac{1-\theta}{\theta} \right).
\end{equation}
For the weak in-chain interaction $\theta \ll 1$. If this is the case it is
necessary to have exponentially small transverse hopping amplitude $t$ in
order to observe non-trivial Q1D effects.

When $\theta > 1$ the effective cut-off momentum $\tilde\Lambda$ is zero.
The quasiparticles are not formed. The system can be viewed as a 
collection of TL bosons weakly coupled by the transverse exchange 
interaction. The possibility of such state was first pointed out by Wen 
\cite{wen}.
It is natural to call such a regime strong coupling. Section V is reserved
for discussion of strong coupling.

\section{Single-electron Green's function}

The calculation of different propagators for Q1D system is an open
question. Our approach allows for easy evaluation of the low-energy part
of Green's functions in the intermediate coupling regime. The high-energy
parts of Q1D Green's functions are
believed to coincide with the Green's functions of TL model. The latter 
have been discussed extensively in the literature.

The Matsubara propagator of the physical electronic field $\psi_{\rm L}$ 
is equal to:
\begin{eqnarray}
{\cal G}_{\rm L}\left(x, {\bf R}_\perp, \tau\right)&=&\frac{1}{\zeta}
\left< {\cal T} \left\{ \Psi_{{\rm L}i}^{\vphantom{\dagger}} 
\left(x,\tau\right) \Psi_{{\rm L}j}^{\dagger} \left(0, 0 \right) \right\} 
\right>_\Psi \times \label{G}\\
&&\left< {\cal T} \left\{ {\rm e}^{-{\rm i}\sqrt{\pi} \left( 
\Theta_{>i}(x,\tau) + \Phi_{>i}(x,\tau) \right)} {\rm e}^{{\rm i}\sqrt{\pi} 
\left( \Theta_{>j}(0, 0) + \Phi_{>j}(0, 0) \right)} \right\} \right>_>,
\nonumber\\
{\bf R}_\perp = {\bf R}_i - {\bf R}_j.
\end{eqnarray}
The notation $\left< \ldots \right>_\Psi$ stands for averaging with respect
to the quasiparticle ground state $\left| 0_\Psi \right>$. Likewise, 
$\left< \ldots \right>_>$ stands for the expectation value with respect to 
$\left| 0_> \right>$ state. 

The bosonic part of this formula can be immediately calculated:
\begin{equation}
\frac{1}{\zeta}\left< {\cal T} \left\{ {\rm e}^{-{\rm i}\sqrt{\pi} \left( 
\Theta_{>i}(x,\tau) + \Phi_{>i}(x,\tau) \right)} {\rm e}^{{\rm i}\sqrt{\pi} 
\left( \Theta_{>j}(0, 0) + \Phi_{>j}(0, 0) \right)} \right\} \right>_>
= \left( \frac{{\cal G}_{\rm L}^{\rm 1d}}{{\tilde{\cal G}}_{\rm L}^{\rm 1d}} 
\right)\delta_{ij} + \zeta^{\theta} (1-\delta_{ij}).
\end{equation}
Here ${\cal G}_{\rm L}^{\rm 1d}$ (${\tilde{\cal G}}_{\rm L}^{\rm 1d}$) 
is the Matsubara Green's function of the
Tomonaga-Luttinger model with the cut-off $\Lambda$ ($\tilde\Lambda$).

Our variational wave function does not take into account correlations
between $\Phi_{>i}(\Theta_{>i})$ and $\Phi_{>j}(\Theta_{>j})$ if $i 
\ne j$. However, the above formula is correct at least for large
$\tau>1/\bar u \tilde \Lambda$ or small frequency $\omega<u\tilde \Lambda$ 
where those correlations are not important. In such a limit the boson 
part of (\ref{G}) is a constant equal to $\zeta^{\theta}$. 

Once the bosonic propagator is found it is necessary to calculate the
quasiparticle Green's function. This can be done with the help of standard
diagrammatic technique.
If we neglect interactions between the quasiparticles the single-electron 
Green's function is equal to:
\begin{equation}
{\cal G}_{\rm L} \left( {\rm i} \omega, p_\|, {\bf p} \right) = 
\frac{\zeta^\theta}{ {\rm i}\omega + v_{\rm F} p_\| -
\tilde\varepsilon^\perp_{{\bf p}} },\label{RGprop}
\end{equation}
where the renormalized transverse kinetic energy is given by:
\begin{equation}
\tilde\varepsilon^\perp_{\bf p} =-2\zeta^\theta
\sum_i t (i) \cos \left({\bf p} \cdot {\bf R}_i\right).\label{Gr}
\end{equation}
This result coincides with the Green's function derived by RG
\cite{bourII,arrigoni}.

Our method allows to improve the above formula for the single-electron
propagator by taking interaction between the quasiparticles into
account. Neglecting (i) symmetry-breaking which becomes important for 
very small frequency only and (ii) the transverse couplings $g_0$ and $\tilde 
g_{2k_{\rm F}}$ (see (\ref{int})) one can identify three second-order
diagrams contributing to the single-quasiparticle self-energy (fig.1). They
are: (a) scattering on the polarization bubble of the same chirality as
the incoming quasiparticle, (b) scattering on the polarization bubble of
the opposite chirality and (c) the vertex correction. The diagrams (a) and
(c) are identical in magnitude and opposite in sign. Thus, (b) is the only
diagram on fig.1 which needs to be evaluated.

First, we calculate the quasiparticle polarization bubble
${\cal P}_{\rm R}$. It equals to:
\begin{equation}
{\cal P}_{\rm R} \left( {\rm i}\Omega, k_\|, {\bf k} \right) =
\int_{q_\| {\bf q}}
\delta \left( v_{\rm F} q_\| + \tilde \varepsilon^\perp_{\bf q} \right)
\frac{v_{\rm F}k_\| + \tilde\varepsilon^\perp_{{\bf q} + {\bf k}} -
\tilde\varepsilon^\perp_{\bf q}}{ {\rm i}\Omega
- v_{\rm F} k_\| - \tilde\varepsilon^\perp_{{\bf q} + {\bf k}} +
\tilde\varepsilon^\perp_{\bf q}},
\end{equation}
where the notation $\int_{q_\| {\bf q}} \ldots = 
(2\pi)^{-3}b^2\int dq_\|d^2 {\bf q} \ldots$ is used. The symbol $b$
denotes the transverse lattice constant.
The self-energy is equal to:
\begin{equation}
{\Sigma}_{\rm L} \left({\rm i}\omega, p_\|, {\bf p} \right) =
-T\sum_\Omega
\int_{k_\| {\bf k}} {\cal G}_{\rm L}\left({\rm i}\omega+{\rm i}
\Omega, p_\| + k_\|, {\bf p} + {\bf k} \right)
{\cal P}_{\rm R} \left({\rm i}\Omega, k_\|, {\bf k} \right).
\end{equation}
After summing over $\Omega$ the following expression for the self-energy is
derived:
\begin{eqnarray}
{\Sigma}_{\rm L} = -g^2 \int_{k_\| {\bf k}}\int_{q_\| {\bf q}}
\left( v_{\rm F} k_\| + \tilde\varepsilon^\perp_ {{\bf k} + {\bf q} }
-\tilde\varepsilon^\perp_{\bf q} \right) \delta\left( v_{\rm F} q_\|
+\tilde\varepsilon^\perp_{\bf q} \right)
\times\\
\frac{ n_{\rm F} \left( v_{\rm F}k_\| +
\tilde\varepsilon^\perp_{{\bf k} + {\bf q}}+\tilde\varepsilon^\perp_{\bf q}
\right)- n_{\rm F} \left( v_{\rm F}(k_\| + p_\|) +
\tilde\varepsilon^\perp_{{\bf k} +{\bf p} } \right) }
{ {\rm i}\omega + 2v_{\rm F}k_\| + v_{\rm F}p_\|
-\tilde\varepsilon^\perp_{\bf q} +
\tilde\varepsilon^\perp_{{\bf k} + {\bf q}} -
\tilde\varepsilon^\perp_{{\bf k} + {\bf p}} } + \nonumber\\
g^2\int_{k_\| {\bf k}}\int_{q_\| {\bf q}}
\frac{ \left( v_{\rm F} k_\| + \tilde\varepsilon^\perp_{{\bf k} + {\bf q}}
-\tilde\varepsilon^\perp_{\bf q} \right) \delta\left( v_{\rm F} q_\|
+\tilde\varepsilon^\perp_{\bf q} \right)
}{
\sinh \left\{ \left( v_{\rm F}k_\| +
\tilde\varepsilon^\perp_{{\bf k} + {\bf q}} +
\tilde\varepsilon^\perp_{\bf q} \right)/T\right\}
\left({\rm i}\omega + 2v_{\rm F}k_\| + v_{\rm F}p_\|
-\tilde\varepsilon^\perp_{\bf q} +
\tilde\varepsilon^\perp_{{\bf k} + {\bf q}} -
\tilde\varepsilon^\perp_{{\bf k} + {\bf p}} \right)}. \nonumber
\end{eqnarray}
When $T=0$ the first term can be further simplified. The Fermi distribution
$n_{\rm F}$ becomes the step-function. In such a situation it is possible to
perform integration over $k_\|$ and $p_\|$ exactly.
The second integral in the above equation appears due to the relation
between the Fermi distribution $n_{\rm F}$ and the Bose distribution
$n_{\rm B}$: $n_{\rm B}(\omega) + n_{\rm F}(\omega) =
1/\sinh(\omega/T)$. At zero temperature this integral vanishes.
In the resultant $T=0$ expression for $\Sigma_{\rm L}$ the
transverse kinetic energy $\tilde\varepsilon^\perp$ always enter in the
combination $\tilde\varepsilon^\perp_{{\bf q} + {\bf k}} +
\tilde\varepsilon^\perp_{{\bf k} + {\bf p}} -
\tilde\varepsilon^\perp_{\bf q}$.
Therefore, it is convenient to introduce the quantity:
\begin{equation}
\nu^\perp (\varepsilon^\perp, {\bf p} ) = \int \frac{d^2
{\bf q} d^2{\bf k}}{(2\pi)^4} b^4
\delta\left(\varepsilon^\perp
-\tilde\varepsilon^\perp_{{\bf q} + {\bf k}} -
\tilde\varepsilon^\perp_{{\bf k} + {\bf p}} +
\tilde\varepsilon^\perp_{\bf q}\right).
\end{equation}
With this definition the self-energy can be compactly written as follows:
\begin{eqnarray}
{\Sigma}_{\rm L}\left(i\omega, p_\|, {\bf p} \right)&=&
\frac{g^2}{8\pi v_{\rm F}} p_\| \label{self-en}\\
&&- \frac{g^2}{16\pi v_{\rm F}^2} \int d\varepsilon^\perp \nu^\perp 
(\varepsilon^\perp, {\bf p} ) \left({\rm i}\omega - v_{\rm F} p_\| - 
\varepsilon^\perp \right) \log\frac{4 \tilde\Lambda^2}{\omega^2 + 
(v_{\rm F} p_\| + \varepsilon^\perp )^2}.\nonumber
\end{eqnarray}
The Green's function of the physical electron is equal to $\zeta^\theta 
({\rm i}\omega+v_{\rm F} p_\| - \tilde\varepsilon^\perp -
\Sigma_{\rm L})^{-1}.$
Note, that the logarithmic divergence of the self-energy, a hallmark of the 
Fermi liquid picture break-down in TL model, is capped in the presence of 
the transverse hopping. This justifies the use of the perturbation theory.

By doing analytical continuation of (\ref{self-en}) it is possible to
calculate the retarded self-energy $\Sigma^{\rm ret}$ whose imaginary part is
the quasiparticle damping:
\begin{equation}
\gamma = - {\rm Im\ \!} \Sigma^{\rm ret} =
\frac{g^2}{8 v_{\rm F}^2} \nu^\perp(-v_{\rm F}p_\|, {\bf p}) \omega^2.
\end{equation}
The transverse density of states can be estimated as $\nu^\perp \propto
1/\tilde t$. This gives us $\gamma \propto (g/v_{\rm F})^2
\omega^2/\tilde t$.
On the mass shell $\omega = -v_{\rm F}p_\| + \tilde \varepsilon_{\bf p}$
the expression for $\gamma$ becomes:
\begin{equation}
\gamma = \frac{g^2}{8} \nu^\perp(-v_{\rm F} p_{\rm F}, {\bf p})
\left(p_\| - p_{\rm F} \right)^2 \propto \frac{g^2}{\tilde t}
\left(p_\| - p_{\rm F} \right)^2 
\end{equation}
where $(p_\| - p_{\rm F} )$ is the distance from a given point $(p_\|,
{\bf p})$ of the Brillouin zone to the Fermi surface $v_{\rm F}
p_{\rm F} = \tilde \varepsilon_{\bf p}$ along $x$ direction.

We need to issue a warning in connection to the accuracy of
$\Sigma_{\rm L}$. It is not correct to think of (\ref{self-en}) as
${\cal O}(g^2)$ expression for {\it the physical electron}
self-energy. Indeed, the physical electron Green's function (\ref{RGprop})
already contains all orders of $g$ entering though the quasiparticle
renormalization $\zeta^\theta$ and renormalized transverse hopping
$\tilde \varepsilon$. It is necessary to remember that our variational
approach is uncontrollable approximation. It lacks a small parameter
controlling the quality of the results. Therefore, it is not clear how
accurate the expression (\ref{self-en}) is.

In Ref. \cite{arrigoni} the self-energy was evaluated numerically for 
the system with infinite transverse dimensions. However, those
calculations are more complicated technically and do not give analytical
answer for the self-energy.

\section{Phase diagram}

In Section II we derived the low-energy effective hamiltonian for
the quasiparticles. Now we will apply the mean field theory to obtain
the phase diagram of the effective hamiltonian. The experimentally observable
phase diagram for the physical electrons coincides exactly with that of the
quasiparticles. To prove this let us calculate
$\left<\psi^\dagger_{{\rm L}i} \psi^{\vphantom{\dagger}}_{{\rm R}i}
\right>$ for $T \ll \bar u \tilde \Lambda$:
\begin{equation}
\left<\psi^\dagger_{{\rm L}i} \psi^{\vphantom{\dagger}}_{{\rm R}i} \right> =
\frac{1}{\zeta} \left< \Psi_{{\rm L}i}^\dagger \Psi_{{\rm
R}i}^{\vphantom{\dagger}} \right>_\Psi \left< {\rm e}^{i\sqrt{\pi}
\left( \Theta_{>i} + \Phi_{>i} \right) }  {\rm e}^{-i\sqrt{\pi} 
\left( \Theta_{>i} - \Phi_{>i} \right) } \right>_> =
\zeta^{{\cal K}-1} \left< \Psi_{{\rm L}i}^\dagger 
\Psi_{{\rm R}i}^{\vphantom{\dagger}} \right>_\Psi.
\end{equation}
The physical CDW order parameter is proportional to the CDW expectation value
of the quasiparticles. Similar formulas can be obtained for other order
parameters. For example, $\left< \psi_{{\rm L}i}^\dagger
\psi_{{\rm R}i}^{{\dagger}} \right> = \zeta^{1/{\cal K}-1}
\left< \Psi_{{\rm L}i}^\dagger \Psi_{{\rm R}i}^{{\dagger}}
\right>_\Psi$ and $\left< \psi_{{\rm L}i}^\dagger
\psi_{{\rm R}j}^{{\dagger}} \right> = \zeta^{\theta}
\left< \Psi_{{\rm L}i}^\dagger \Psi_{{\rm R}j}^{{\dagger}}
\right>_\Psi$.
Therefore, we can determine the phase diagram of (\ref{H}) by mapping 
the phases of the hamiltonian (\ref{Heff}).

We consider four order parameters. One is the charge-density wave:
\begin{eqnarray}
\hat \rho_{2k_{\rm F}i} = \Psi_{{\rm L}i}^\dagger \Psi_{{\rm
R}i}^{\vphantom{\dagger}}
\end{eqnarray}
and there are three types of the superconducting order:
\begin{eqnarray}
\hat\Delta_{\pm ij} = \frac{1}{2} 
\left(\Psi_{{\rm L}i}^\dagger \Psi_{{\rm R}j}^\dagger \pm
\Psi_{{\rm L}j}^\dagger \Psi_{{\rm R}i}^\dagger \right), \\
\hat\Delta_{0i} = \Psi_{{\rm L}i}^\dagger \Psi_{{\rm R}i}^\dagger.
\end{eqnarray}
The in-chain potential energy can be re-written in terms of $\hat\rho$ and
$\hat \Delta_0$ in the following manner:
\begin{equation}
g\Psi_{{\rm L}i}^\dagger \Psi_{{\rm L}i}^{\vphantom{\dagger}}
\Psi_{{\rm R}i}^\dagger \Psi_{{\rm R}i}^{\vphantom{\dagger}} = - g
\hat\rho_{2k_{\rm F}i}^{\vphantom{\dagger}} \hat\rho_{2k_{\rm F}i}^\dagger
= g\hat\Delta_{0i}^\dagger \hat\Delta_{0i}^{\vphantom{\dagger}}.
\end{equation}
The exchange interaction can be expressed as:
\begin{equation}
\tilde g_{2k_{\rm F}} \left( \Psi^\dagger_{{\rm L}i}
\Psi^{\vphantom{\dagger}}_{{\rm R}i} \Psi^\dagger_{{\rm R}j}
\Psi^{\vphantom{\dagger}}_{{\rm L}j} + {\rm h.c.}\right)
=\tilde g_{2k_{\rm F}} \left(
\hat\rho_{2k_{\rm F}i}^{\vphantom{\dagger}} \hat\rho_{2k_{\rm F}j}^\dagger
+ {\rm h.c.}\right)
= 2\tilde g_{2k_{\rm F}} \left( \hat\Delta_{-ij}^{\vphantom{\dagger}}
\hat\Delta_{-ij}^\dagger - \hat\Delta_{+ij}^{\vphantom{\dagger}}
\hat\Delta_{+ij}^\dagger \right).\label{ex}
\end{equation}
Finally, a part of the transverse forward scattering which describes the
interaction between the fermions of different chiralities is equal to:
\begin{equation}
g_0 \left( \Psi^\dagger_{{\rm L}i}
\Psi^{\vphantom{\dagger}}_{{\rm L}i} \Psi^\dagger_{{\rm R}j}
\Psi^{\vphantom{\dagger}}_{{\rm R}j} + 
\Psi^\dagger_{{\rm R}i}
\Psi^{\vphantom{\dagger}}_{{\rm R}i} \Psi^\dagger_{{\rm L}j}
\Psi^{\vphantom{\dagger}}_{{\rm L}j} \right) = 
2g_0 \left( \hat \Delta_{+ij}^{\vphantom{\dagger}} 
\hat \Delta_{+ij}^\dagger + \hat \Delta_{-ij}^{\vphantom{\dagger}} 
\hat \Delta_{-ij}^\dagger \right).\label{fs}
\end{equation}
The part of the forward scattering which accounts for the interaction 
between the fermions of the same chirality cannot be expressed in terms of 
these four order parameters.

The effective coupling for CDW is always bigger then the effective
coupling for the superconducting order parameter $\hat \Delta_+$:
\begin{eqnarray}
g_{\rm CDW} > g_{\rm sc},\ {\rm where}\\
g_{\rm CDW} = g+z_\perp \tilde g_{2k_{\rm F}},\\
g_{\rm sc}= \tilde g_{2k_{\rm F}} - g_0 =
 \zeta^{2{\cal K} - 2} g_{2k_{\rm F}} - g_0,
\label{gsc}
\end{eqnarray}
and $z_\perp$ is the coordination number for a chain.  Thus, at $T=0$ for
the perfect nesting the system is always in CDW phase with $\hat \Delta_+$
order parameter phase being meta-stable ($g_{\rm sc}>0$) or unstable 
($g_{\rm sc}<0$). Other order parameters, $\hat \Delta_0$ and
$\hat \Delta_-$, are unstable. 

When the external pressure is applied the amplitude $t_2$ for hopping to
the next-to-nearest chain begins to grow and spoils the Fermi surface 
nesting. This undermines stability of CDW and drives the transition 
temperature to zero \cite{sdw}. Indeed, in the latter reference the 
following simple estimate for the density wave susceptibility was obtained:
\begin{equation}
\chi \propto \frac{1}{2\pi v_{\rm F}}\times \cases{
\log\left( 2v_{\rm F}\tilde\Lambda/T \right),& if $T>\tilde t_2=\zeta^\theta
t_2$\cr
\log\left( 2v_{\rm F}\tilde\Lambda/ \tilde t_2\right),& if $T<\tilde 
t_2=\zeta^\theta t_2$,}
\label{chi}
\end{equation}
The CDW transition temperature is derived by equating $(g + z_\perp 
\tilde g_{2k_{\rm F}})\chi$ and unity. For $\tilde t_2 = 0$ it is:
\begin{equation}
T_{\rm CDW}^{(0)} \propto v_{\rm F}\tilde\Lambda \exp\left(-2\pi v_{\rm F}/
\left( g + z_\perp \tilde g_{2k_{\rm F}}\right)\right).\label{T0}
\end{equation}
If $\tilde t_{2} > 0$ the transition temperature $T_{\rm CDW}$ becomes 
smaller then $T_{\rm CDW}^{(0)}$. It vanishes when
$\tilde t_{2} \propto T_{\rm CDW}^{(0)}$.
That is, exponentially small $\tilde t_2$ is enough to destroy CDW.

What happens after CDW is destroyed depends on the sign of $g_{\rm sc}$.
If $g_{\rm sc} > 0$ the ground
state is superconducting. Otherwise, it is the Fermi liquid. We can 
perform the same type of analysis we did above for CDW.
The superconductivity is rather insensitive to the nesting properties of
the Fermi surface. The susceptibility for $\hat \Delta_+$  is equal to
$(1/2\pi \alpha v_{\rm F})
{\log\left( 2v_{\rm F} \tilde\Lambda / T\right)}$, where $\alpha$ is a
constant of order of unity. The critical temperature is found to be:
\begin{eqnarray}
T_c \propto v_{\rm F} \tilde\Lambda \exp \left( -2\pi \alpha v_{\rm F} /
g_{\rm sc} \right),\label{crit}
\end{eqnarray}
if $g_{\rm sc}>0$. Even when $g_{2k_{\rm F}} < g_0$ the 
effective coupling $g_{\rm sc}$ may be positive provided that the in-chain
interaction is repulsive (${\cal K}<1$) and the electron hopping anisotropy 
parameter $(\bar u \Lambda / t)$ is big:
\begin{equation}
\zeta^{2{\cal K} - 2} >
\frac{g_0}{g_{2k_{\rm F}}} \Leftrightarrow
\left( \frac{\bar u \Lambda}{t} \right)^{(2-2{\cal K})/(1-\theta)} >
\frac{g_0}{g_{2k_{\rm F}}} > 1.\label{stable}
\end{equation}
For the system in the intermediate coupling regime this condition is likely
to be satisfied.

It is interesting to note that the external pressure detriments not only
CDW but the superconductivity as well. Under pressure the anisotropy
parameter $(\bar u \Lambda / t)$ decays. The superconducting 
transition temperature gets smaller as the anisotropy decreases. At pressure 
higher then some critical value the condition (\ref{stable}) is no longer
satisfied. In this region the superconductivity is unstable and the ground
state is the Fermi liquid.

The qualitative phase diagram is presented on the fig.2. It shares two
remarkable features with the phase diagram of the organic Q1D 
superconductors \cite{book}: (i) the density wave phase and 
superconductivity have common boundary; (ii) the superconducting transition 
temperature vanishes at high pressure.

Our order parameter $\hat\Delta_+$ deviates from the more common version
$\hat\Delta_0$. The order parameter $\hat\Delta_+$ was proposed
quite some time ago \cite{emery}. Recently, this suggestion found further 
support in the renormalization group calculations of Ref. \cite{dup}. The 
advantage of $\hat\Delta_+$ stems from the fact that by having two 
electrons of a Cooper pair on different chains we avoid increasing 
in-chain potential energy.

The origin of the superconducting phase in our system is an interesting 
question worth discussing in more details. In conventional BCS model the
superconductivity is stable because it minimizes the potential energy of
the electron-electron interaction. We can make this claim rigorous by
considering the following derivation. BCS hamiltonian density
\begin{equation}
{\cal H}^{\rm BCS} = {\cal T} + {\cal V} =
\sum_\sigma \psi^\dagger_\sigma \left( \frac{\hat p^2}{2m} - \mu \right)
\psi^{\vphantom{\dagger}}_\sigma 
- g \psi^\dagger_\uparrow \psi^\dagger_\downarrow
\psi^{\vphantom{\dagger}}_\downarrow \psi^{\vphantom{\dagger}}_\uparrow
\end{equation}
consists of two term: kinetic energy density $\cal T$ and potential energy
density $\cal V$. At zero temperature the superconducting state energy
density ${\cal E}_{\rm sc}=\left< {\cal H} \right>_{\rm sc}$ is smaller
then the normal energy density
${\cal E}_{\rm n}=\left< {\cal H} \right>_{\rm n}$. This condensation
energy density
\begin{eqnarray}
{\cal E}_{\rm c} = {\cal E}_{\rm n} - {\cal E}_{\rm sc} \propto \nu T^2_c,
\\
\nu = \pi^{-2} m^2 v_{\rm F},\ v_{\rm F} = \sqrt{2\mu/m},
\end{eqnarray}
is entirely due to depletion of interaction in the superconducting
state: 
\begin{equation}
\left< {\cal V} \right>_{\rm n}-\left< {\cal V} \right>_{\rm sc} > 0.
\label{V}
\end{equation}
As for the kinetic energy it grows in the superconducting state:
\begin{equation}
\left< {\cal T} \right>_{\rm n}-\left< {\cal T} \right>_{\rm sc} <0.
\label{T}
\end{equation}
To prove this we will use Feynman formula which allows to calculate
the ground state expectation value of any term $c {\cal O}$ of the
hamiltonian density:
\begin{equation}
c\langle {\cal O} \rangle = c\frac{\partial {\cal E}_{\rm gs}}{\partial c},
\end{equation}
where ${\cal E}_{\rm gs}$ is the ground state energy. Therefore:
\begin{eqnarray}
\left< {\cal V} \right>_{\rm n}-\left< {\cal V} \right>_{\rm sc} =
g\frac{\partial}{\partial g} {\cal E}_{\rm c},\\
\left< {\cal T} \right>_{\rm n}-\left< {\cal T} \right>_{\rm sc} =
m^{-1}\frac{\partial}{\partial {m^{-1}}} {\cal E}_{\rm c}, \\
{\cal E}_{\rm c} \propto \omega_{\rm D}^2 \mu^{1/2} m^{3/2} \exp
( - \alpha \mu^{-1/2} m^{-3/2} g^{-1}),
\end{eqnarray}
where $\omega_{\rm D}$ is Debye frequency and $\alpha>0$ is a constant of
order unity. The inequalities (\ref{V}) and (\ref{T}) immediately follow
from the expressions above. These inequalities mean that it is the
electron-electron attraction which triggers BCS superconductivity. This
fact is a very well known fact of the superconductivity mean-field theory.

However, in the system with strong repulsion, such as Q1D or high-$T_c$
materials, it is difficult to construct a mean-field superconducting phase
which lowers the interaction energy. Our model for which we develop the
consistent many-body approach can be used to discuss this issue beyond the
mean-field approximation.

For our model it is easy to determine that the transverse forward
scattering energy is increased and the exchange energy is decreased by the
superconductivity. This result is a direct consequence of (\ref{fs}) and
(\ref{ex}). 

Contributions of other terms can be found with the help of Feynman formula.
The condensation energy density is of the order of
$-T^2_c/v_{\rm F}$. Thus, differentiating the
critical temperature (\ref{crit}) with respect to some coupling constant of
(\ref{H}) we can determine how a ground state energy contribution of a
given term is modified by presence of the superconductivity. A derivative
of the critical temperature with respect to a parameter $x$ is equal to:
\begin{equation}
\frac{\partial}{\partial x} T_c = T_c \left(\frac{\partial}{\partial x}
\log\tilde\Lambda + \frac{v_{\rm F}}{g_{\rm sc}} \frac{\partial}{\partial x}
\log g_{\rm sc} \right) \approx 
T_c \frac{v_{\rm F}}{g_{\rm sc}} \frac{\partial}{\partial x}\log g_{\rm sc},
\end{equation}
provided that $g_{\rm sc} \ll v_{\rm F}$. Combining this result with 
(\ref{gsc}) we conclude that in the superconducting state the transverse
hopping energy is higher:
\begin{eqnarray}
\left< -t\sum_{p\langle i,j \rangle} \left(
 \psi^\dagger_{pi} \psi^{\vphantom{\dagger}}_{pj} + {\rm h.c.} \right)
\right>_{\rm n} -
\left< -t\sum_{p\langle i,j \rangle} \left(
 \psi^\dagger_{pi} \psi^{\vphantom{\dagger}}_{pj} + {\rm h.c.} \right)
\right>_{\rm sc}\propto \\
t \frac{T_c^2}{g_{\rm sc}}
\frac{\partial}{\partial t} \log \left( g_{2k_{\rm F}}
\left(\frac{t}{\bar u \Lambda}
\right)^\frac{2{\cal K} - 2}{1-\theta} - g_0 \right) < 0\nonumber
\end{eqnarray}
and the in-chain potential energy is lower then in the normal state:
\begin{eqnarray}
\left<g\psi_{{\rm L}i}^\dagger \psi_{{\rm L}i}^{\vphantom{\dagger}}
\psi_{{\rm R}i}^\dagger \psi_{{\rm R}i}^{\vphantom{\dagger}}
\right>_{\rm n}-
\left<g\psi_{{\rm L}i}^\dagger \psi_{{\rm L}i}^{\vphantom{\dagger}}
\psi_{{\rm R}i}^\dagger \psi_{{\rm R}i}^{\vphantom{\dagger}}
\right>_{\rm sc}
\propto g \frac{T_c^2\tilde g_{2k_{\rm F}}}{g_{\rm sc}^2} \log \left(
\frac{t}{\bar u \Lambda} \right) \frac{\partial}{\partial g} 
\left( \frac{2{\cal K} - 2}{1-\theta}\right) > 0,
\end{eqnarray}
since both $\log ({t}/{\bar u \Lambda} )$ and the derivative with respect
to $g$ are negative.

We have proven that in our case the superconductivity is triggered by the
electron-electron repulsion. This result is quite unexpected. It is has a
many-body nature and cannot be obtained within a mean-field theory for
the hamiltonian (\ref{H}). This mechanism of superconductivity is very
similar to the Kohn-Luttinger proposal. Classical Kohn-Luttinger
mechanism predicts extremely low critical temperature. In our case, however,
the effective coupling constant $g_{\rm sc}$ is a non-analytical function
of the bare parameters. As a consequence, our transition temperature
(\ref{crit}) does not have to be small.

\section{Strong coupling regime}

We have seen above that if $\theta>1$ then $\tilde\Lambda$ is zero. This
means that quasiparticles are not formed and it is more convenient to treat
the system in terms of TL boson only. The bosonized hamiltonian (\ref{H})
has the form:
\begin{eqnarray}
{\cal H}^{\rm 1d}_i = \frac{u}{2} \left( {\cal K}\left(\nabla \Theta_i
\right)^2 + {\cal K}^{-1}\left(\nabla \Phi_i \right)^2\right),\\
{\cal H}_{ij}^\perp = \frac{g_{2k_{\rm F}}}{(2\pi)^2} \cos \sqrt{4\pi} 
\left( \Phi_i - \Phi_j \right).\label{exch}
\end{eqnarray}
In this formula both the transverse hopping term which is irrelevant in 
RG sense and the forward scattering term which is marginal are omitted. 
Their effect is small as compared with that of the strongly relevant 
exchange interaction, eq. (\ref{exch}).

The relevance of the exchange interaction indicates that at low temperature
the system freezes into a state with the finite expectation value $\langle
\Phi_i \rangle \ne 0$. This phase is CDW. It can be easily proved by
bosonizing CDW order parameter: $\psi_{{\rm L}i}^\dagger\psi_{{\rm
R}i}^{\vphantom{\dagger}} \propto (2\pi a)^{-1} \exp({\rm
i}\sqrt{4\pi}\Phi_i)$. The finite expectation value of the field $\Phi$
is inherited by CDW order parameter.

We describe this regime with the help of our variational wave function.
Since $\tilde \Lambda = 0$ one can write the wave function in terms of TL
bosonic field only:
\begin{equation}
\left| {\rm Var} \right> = \prod_{k>0,i} \frac{1}{\sqrt{2\pi \sigma_k^2}}
\exp \left\{ - \left|\Phi_{ik} \right|^2 / 4\sigma_k^2 \right\}.
\label{vari2}
\end{equation}
This expression is a slight generalization of (\ref{vari}): in the latter
equation the parameters $\sigma_k^2 = {\cal K}/4|k|$. Here we do not fix
$\sigma_k^2$. Instead, they will be determined variationally. Variational
energy is:
\begin{equation}
E^{\rm V}/(LN_\perp) = u \int_0^{\Lambda} \frac{dk}{2\pi} \left( 
\frac{\cal K}{ 8\sigma^2_k} + \frac{2k^2 \sigma_k^2}{\cal K} \right)
-g_{2k_{\rm F}}\Lambda^2 \exp \left\{ -8 \int_0^\Lambda dk\sigma_k^2
\right\}.
\end{equation}
Minimizing this energy with respect to $\sigma_k^2$ we find:
\begin{eqnarray}
\sigma_k^2 = \frac{{\cal K}u}{4\sqrt{u^2k^2 + \Delta_{\rm CDW}^2}},\\
\Delta_{\rm CDW}^2 = 8\pi g_{2k_{\rm F}}{\cal K}u \Lambda^2
\exp \left\{ -2{\cal K}\int_0^{u\Lambda} \frac{d\varepsilon}
{\sqrt{\varepsilon^2 + \Delta_{\rm CDW}^2}} \right\}
\propto g_{2k_{\rm F}}u \Lambda^2 \left( \frac{\Delta_{\rm CDW}}
{u\Lambda} \right)^{2{\cal K}}.
\end{eqnarray}
The quantity $\Delta_{\rm CDW}$ has the meaning of the excitation gap due to
CDW order. This gap, together with the transition temperature, can be found
by solving the last equation:
\begin{equation}
T_{\rm CDW} \propto \Delta_{\rm CDW} \propto u\Lambda
\left( \frac{g_{2k_{\rm F}}}{u} \right) ^{1/(2-2{\cal K})}.\label{Tcdw}
\end{equation}
The variational energy is:
\begin{equation}
E^{\rm V}/(LN_\perp) \propto \Delta_{\rm CDW}^2/u.\label{Ecdw}
\end{equation}
These results are correct when ${\cal H}^\perp$ couples only those chains
which are nearest neighbors. The next-to-nearest neighbor coupling
frustrates CDW phase. We will not discuss the effect of the frustration in
this paper.

Finally, let us discuss cross-over from strong to intermediate coupling
regime. Such cross-over occurs when the intermediate coupling Fermi liquid
energy, eq. (\ref{EVs}), becomes equal to the strong coupling CDW energy,
eq.(\ref{Ecdw}):
\begin{equation}
{\tilde t}^2/v_{\rm F} \propto \Delta_{\rm CDW}^2/u\ \ {\rm or}\ 
\left(\frac{t}{\bar u \Lambda}\right)^{1/(1-\theta)}
\propto
\left( \frac{g_{2k_{\rm F}}}{u}\right)^{1/(2-2{\cal K})}.
\end{equation}
This equation defines $\theta_c (t, g_{2k_{\rm F}}) < 1$ at which the
cross-over takes place. At small $\theta<\theta_c$ the system behaves as
the Fermi liquid whose properties we discussed in the previous sections.
When $\theta > \theta_c$ the expression (\ref{L8})
is no longer applicable: the necessary requirement for smallness of
the energy associated with symmetry breaking is violated. The expression
(\ref{Ecdw}) has to be used instead.

Fig.3 shows how the strong coupling regime at big $\theta$ is replaced by
the intermediate coupling regime at smaller $\theta$. The transition
temperature of CDW, eq.(\ref{Tcdw}), drops sharply and becomes exponentially
small, eq.(\ref{T0}), as $\theta$ gets smaller then $\theta_c$. This
diagram was discussed in \cite{boies} for a similar model.

\section{Conclusions}

We propose in this paper the variational wave function for Q1D system. Our
procedure key ingredient is the splitting of TL bosons into high-momentum
and low-momentum modes. While high-momentum modes are in their ground state
the low-momentum modes form quasiparticles which delocalize in the
transverse directions.

Our method can be viewed as a variational implementation of the lowest
order RG scaling near TL liquid fixed point. When the transverse hopping
amplitude becomes of the order of $u\Lambda$ the scaling must be stopped.
The renormalized hamiltonian should be treated as the hamiltonian for 
the quasiparticles. 

Our method gives us a possibility of computing different 
Green's functions beyond RG using standard diagrammatic technique. 
As an example we calculated the lowest order self-energy for the 
one-particle propagator.

Depending on the strength of the in-chain interaction and the anisotropy 
the system may be in one of three regimes: strong, intermediate or weak
coupling. In the strong coupling regime quasiparticles are not formed and
the system is better described in terms of TL bosons. In weak and
intermediate coupling regime the low-lying degrees of freedom are
quasiparticles.  The ground state of these fermions may
be either Fermi liquid, the superconductivity or CDW. The phase diagram
of our Q1D model looks very similar to that of the organic Q1D
superconductors.

Unlike classical BCS superconducting phase, the one in our model is
stabilized without any attraction between the electrons. It is similar to
Kohn-Luttinger superconductivity. However, our effective coupling constant
is bigger then that of Kohn-Luttinger. This guarantees that the critical
temperature in our model is not unacceptably small.

\begin{figure} [!t]
\centering
\leavevmode
\epsfxsize=8cm
\epsfysize=8cm
\epsfbox[18 -100 592 718] {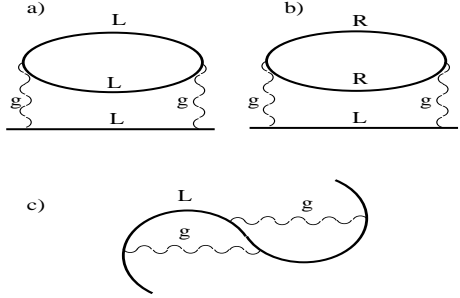}
\caption[]
{\label{fig5} 
Lowest order contribution to the self-energy of the quasiparticle.
}
\end{figure}
\begin{figure} [!t]
\centering
\leavevmode
\epsfxsize=8cm
\epsfysize=8cm
\epsfbox[18 -100 592 718] {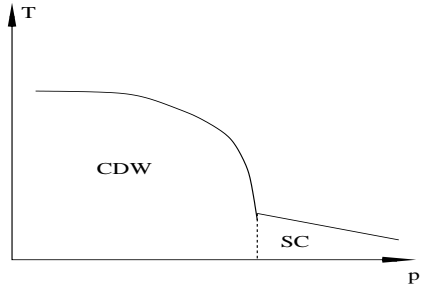}
\caption[]
{\label{fig1} 
Qualitative phase diagram of our model. Solid lines show second-order
phase transitions into CDW and the superconducting phase. Dashed line shows
the first-order transtion between CDW and the superconductivity.
}
\end{figure}

\begin{figure} [!t]
\centering
\leavevmode
\epsfxsize=8cm
\epsfysize=8cm
\epsfbox[18 -100 592 718] {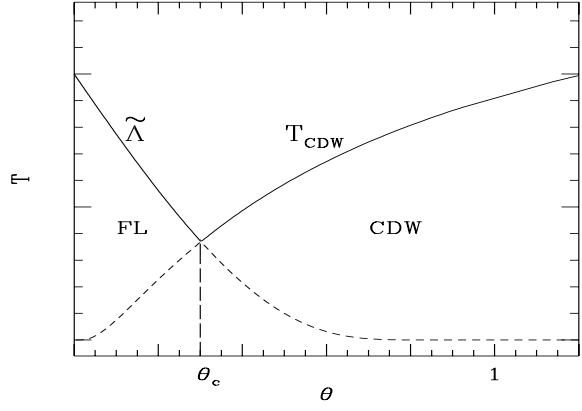}
\caption[]
{\label{fig2} 
The energy scale assosiated with transverse hopping $\tilde \Lambda$
decreases when $\theta$ grows. The CDW transition temperature
$T_{\rm CDW}$ increases
as $\theta$ grows. At $\theta_c$ where both energy scales are of the same
order the cross-over from the intermediate to strong coupling occurs.
}
\end{figure}

\end{document}